# Detection of Quantum Noise from an Electrically-Driven Two-Level System

Richard Deblock[1,2], Eugen Onac[1], Leonid Gurevich[1], Leo P. Kouwenhoven[1]

**Quantum mechanics can strongly influence the noise properties of mesoscopic devices. To probe this effect we have measured the current fluctuations at high-frequency (5-90 GHz) using a superconductor-insulator-superconductor tunnel junction as an on-chip spectrum analyser. By coupling this frequency-resolved noise detector to a quantum device we can measure the high-frequency, non-symmetrized noise as demonstrated for a Josephson junction. The same scheme is used to detect the current fluctuations arising from coherent charge oscillations in a two-level system, a superconducting charge qubit. A narrow band peak is observed in the spectral noise density at the frequency of the coherent charge oscillations.**

[1] Department of Nanoscience and ERATO Mesoscopic Correlation Project, Delft University of Technology, POBox 5046, 2600 GA Delft, Netherlands
[2] Present address: Laboratoire de Physique des Solides, associé au CNRS, Université Paris-Sud, Bâtiment 510, 91 405 Orsay Cedex, France

Noise, i.e. current fluctuations, has proved to be a powerful tool to probe mesoscopic devices (*1*). At high frequency it can bear strong signature of the dynamics resulting from quantum mechanics. One of the simplest system to study this effect is a two-level system (TLS) with two coupled quantum states, |0> and |1>, which can form a coherent superposition, $\alpha$ |0> + $\beta$ |1>, with $\alpha$ and $\beta$ complex numbers (see Fig. 1A). If this TLS is forced into state |0> at time equal to zero, the probability, $P_{|1>} = |\beta|^2$, to find the system in state |1> oscillates in time with a frequency determined by the coupling strength. This prediction (*2*) has recently attracted much interest in the context of quantum computation where TLS form the physical realizations of the qubit building blocks. To determine the state of the qubit some detection mechanism is needed. In the case of solid state devices the qubit state is often measured by means of the value of an electrical current (*3*). We are interested in the fluctuations in the read-out current and how these are affected by the oscillating time-evolution of a qubit.

The central idea is illustrated in Fig. 1B and C. Suppose an electron hops on a TLS and initially occupies state |0>. Due to the coupling between the two states (dashed arrow) $P_{|1>}$ starts to oscillate. The electron can leave the TLS towards the right only when $P_{|1>}$ is high. A new electron then repeats the cycle. Thus the outgoing current consists of charge injections that preferentially occur near odd integer times half the oscillation period after the previous tunneling event (see Fig. 1C). The fluctuations in the current still bear this non-stochastic noise and instead of the usual white noise spectrum, a narrow band peak is expected at the frequency determined by the coupling strength.

The idea above is very general and theoretical predictions on narrow band noise exist for Bloch oscillations in a double quantum well (*4*), charge oscillations in superconducting (*5*) and semiconducting qubits (*6*), and electron spin resonance oscillations (*7*). The experimental detection is difficult as the frequency, *f*, of the coherent oscillations is typically in the GHz range in order to fulfil the condition $hf >> k_BT$, with $k_BT$ the thermal energy (*8*). We report a detection scheme from which we obtain the frequency-resolved spectral density of current noise in the range of 5 to 90 GHz (*9*).

Our detection scheme follows the ideas of Refs. (*10,11*) : a quantum device is coupled on-chip to a detector that converts the high-frequency noise signal into a DC current. The on-chip coupling provides a large frequency bandwidth (~100 GHz) whereas the conversion to DC allows standard amplification of the signal (*12*). Our detector is a superconductor-insulator-superconductor (SIS) tunnel junction, known to be a sensitive microwave detector and well established in astronomy measurements (*13*). For low voltage bias, the gap, $\Delta$, in the density of states prohibits tunnelling of quasi-particles (left inset of Fig. 2B). However, for a bias $V_{SIS}$, the absorption of a photon of energy $\hbar\omega$ that exceeds ($2\Delta$-$eV_{SIS}$), can assist tunnelling (right inset of Fig. 2B). This photon-assisted tunneling (PAT) current carries information on the number and the frequency of photons reaching the detector (*14*).

To validate our noise detection we have first measured on a Josephson junction (Jj) for which the high-frequency fluctuations are well known (Fig. 2A) (*15*). A Josephson junction, biased such that $|eV_{Jj}| < 2\Delta$, generates an AC current of frequency $f_{Jj} = 2eV_{Jj}/h$ (*16,17*). The AC current fluctuations are capacitively coupled to the detector side where voltage fluctuations build up across the SIS junction. These electromagnetic fluctuations form microwave photons, which can be absorbed by tunneling quasi-particles. The measured PAT current through the SIS detector is shown in Fig. 2B (black solid curve). A clear step in the current is seen at $V_{SIS}$ = 285 µV. Roughly speaking the derivative of this curve is a measure of the noise spectral density (see below). Thus the noise of a Josephson junction is indeed found to be narrow band and in this case centered around ($2\Delta$–$eV_{SIS}$) = 32.5 GHz consistent with the expected frequency $f_{Jj}$ = 33.8 GHz for $V_{Jj}$ = 70 µV.

For a quantitative description we consider an SIS junction subject to current fluctuations. The PAT current for a bias $eV_{SIS} < 2\Delta$ is given by (*18*):

$$I_{PAT}(V_{SIS}) = \int_{0}^{+\infty} d\omega \left(\frac{e}{\hbar\omega}\right)^2 |Z(\omega)|^2 S_I(-\omega) I_{SIS}(V_{SIS} + \hbar\omega/e) \qquad (1)$$

with $I_{SIS}(V_{SIS})$ the SIS current without noise, $Z(\omega)$ the transimpedance ($Z(\omega) = \{S_{V,SIS}(\omega)/S_I(\omega)\}^{1/2}$, i.e. voltage fluctuations at the detector divided by current fluctuations from the source). We emphasize that here the spectral density $S_I(\omega)$ corresponds to a non-symmetrized noise correlator (*10,11,19,20*). Our SIS detector measures absorption of photons, which were emitted from the Josephson junction. Because the SIS detector itself is virtually noiseless for $eV_{SIS} < 2\Delta$, no emission from the detector occurs and thus no absorption takes place in the Josephson junction. Under these conditions we only measure the spectral density at (as commonly defined) negative frequencies, $S_I(-\omega)$ (*21*).

Equation 1 has been used to calculate the dashed curve in Fig. 2B assuming a time dependent current $I(t)=I_C \sin(2\pi f_{Jj} t)$ (or equivalently two delta-function noise peaks at $+f_{Jj}$ and $-f_{Jj}$). Using this AC Josephson current and formula of ref. 18 we recover the formula for the PAT current (formula 3.4 of Ref. 14). The value of $Z(\omega)$ at $\omega= 2\pi f_{Jj}$ has been used as a fitting parameter in order to obtain good agreement with the experimental curve. We have repeated this procedure for many different frequencies and thus obtained $Z(\omega)$ from 10 to 80 GHz (see Fig. 3B). Note that in this calculation only the noise peak at $-f_{Jj}$ leads to a significant contribution to the PAT current (for $eV_{SIS} < 2\Delta$).

Biasing the Josephson junction such that $eV_{Jj} > 2\Delta$, turns on quasi-particle tunneling. This tunneling is stochastic and gives rise to white shot-noise (*1*). The SIS detector current (curve 2 of Fig. 2B) now appears very different from the narrow-band noise in curve 1. The dashed-dotted curve is calculated without any fit parameter. We have simply inserted $Z(\omega)$ [i.e. the red line in Fig. 3B] and the shot noise value, $S_I(\omega)= eI_{Jj}$ (*22*), in Eqn. 1 and obtain good agreement with the experiment. It is again important to note that $S_I(\omega) = eI_{Jj}$ corresponds to the non-symmetrized noise value (the symmetrized noise is the Poissonian value equal to $2eI_{Jj}$). The fact that only $eI_{Jj}$ fits our data, demonstrates the first observation of non-symmetrized noise in electrical conductors.

Figure 3A is a plot of the detector current versus $V_{Jj}$. The narrow-band AC Josephson noise and the white quasi-particle noise are clearly distinguishable with a transition at $|eV_{Jj}|= 2\Delta$. Fig. 3C shows that the quasi-particle noise depends linearly on $I_{Jj}$ confirming the white shot-noise character of this regime. We find that the resolution of this noise measurement is 80 fA$^2$/Hz, corresponding to an equivalent noise temperature of 3 mK on a 1 k$\Omega$ resistor.

To demonstrate narrow-band noise from an electrically-driven qubit, we have chosen the Cooper Pair Box (CPB) as a physical realization of the TLS. The CPB is fabricated with the same aluminum technology as the SIS detector and therefore easy to integrate on-chip. The two-levels, |0> and |1>, correspond to $N$ and $N+1$ Cooper pairs in the box, which is controlled by the gate voltage $V_g$ (Fig. 4B, inset). The two levels are coupled by the Josephson energy, $E_J$, as already illustrated in Fig. 1A (*3,23*). The coherent charge oscillation corresponds to one extra Cooper pair tunnelling on and off the box. When $P_{|1\rangle}$ is high a sudden decay to the $|0\rangle$ state can take place by quasi-particle tunneling out of the qubit. The resulting current is expected to have narrow-band noise around a frequency $f=\sqrt{(4E_C(Q/e-1))^2+E_J^2}/h$ (*5*), which describes the energy difference between the two-levels in Fig. 1A. $E_C = e^2/2C_\Sigma$ is the charging energy, with $C_\Sigma$ the total capacitance of the island, and $Q = C_g V_g$ is the charge induced on the box. Because the decay is a stochastic process occurring around odd-multiples times half the oscillation period, the narrow-band noise is not a delta-peak as in the case of the AC Josephson effect. Instead, a broad peak is expected on top of a white shot-noise background (*5*). The sudden quasi-particle decay is realized for bias and gate voltages near, the so-called Josephson-QuasiParticle (JQP) peak (*3*). The average number of coherent charge oscillations is determined by the ratio $E_J/\Gamma$ where $\Gamma$ is the decay rate for the two quasi-particles (*24*). In our device the Josephson junction has a SQUID geometry allowing to tune $E_J$. Consequently, we can explore both the coherent ($E_J > \Gamma$) and incoherent ($E_J < \Gamma$) regime.

The CPB (Fig. 4, B and C) is coupled to an SIS detector with the same on-chip circuitry as in Fig. 2 (fig S1). First the CPB is characterized independently and the JQP peak identified (Fig. 4A, inset). Fig. 4A shows measurements of the PAT current through the SIS detector for different CPB gate voltages. The PAT current is rather high on the left side ($Q/e < 1$) of the JQP peak and small on the right side ($Q/e > 1$). This is attributed to the emission character of the left side of the JQP peak versus absorption on the right (*25*). On the absorption side ($Q/e > 1$) the small PAT signal is attributed to tunnelling processes not related to the coherent dynamics of the CPB, leading to a background PAT current. Indeed since no energy is available flowing towards the CPB the spectral density of the current fluctuations on the absorption side is virtually zero.

On the left side we observe more high-frequency components when moving away from the JQP peak center. To extract the noise component related to the JQP process we subtract the small PAT current at high $Q$ (Fig. 4B). We determine then the dominant frequency component from the $V_{SIS}$-value where PAT becomes visible (we checked the validity of this determination of the dominant frequency for the AC Josephson effect) . The $V_{SIS}$-values converted to frequencies are plotted versus $Q$ (Fig. 4C). The dominant frequency dependence on the charge on the CPB, can be fitted by the energy difference between the two-levels of the CPB (solid curves in Fig. 4C). We obtain a charging energy slightly higher than expected. The dominant frequency value for $Q/e = 1$ is consistent with the value of $E_J$ in this sample ($E_J$ = 50 μeV = 12 GHz for maximum coupling (*26*)). Changing $E_J$ (by means of the flux through the SQUID) changes the dominant frequency for $Q/e = 1$ as shown for the red data points. For small values of $E_J$ the PAT signal becomes very weak and has a shot-noise shape similar to the PAT curves for $Q/e > 1$ (Fig. 4A). Indeed, in this incoherent regime ($\Gamma > E_J$) a dominant frequency from narrow-band noise is not expected.

We have demonstrated narrow-band high-frequency detection of non-symmetrized noise. The quantum noise from a charge qubit shows a peak at the frequency of the coherent charge oscillation. The SIS detector is operated as an on-chip spectrum analyser and is applicable for correlation measurements on a wide range of electronic quantum devices.

27. We thank Y. Nakamura, K. Harmans, P. Hadley, Y. Nazarov, H. Mooij, D. Bragett and Y. Blanter for discussions. We acknowledge the technical assistance of R. Schouten, B. van der Enden and M. van Oossanen. Financial support is obtained from the Dutch Organisation for Fundamental Research (FOM), and from ARO.

**Figure 1.** (**A**) Energy diagram for a two-level system with parameters for a superconducting charge qubit of charging energy $E_C$. $Q$ is the qubit charge. The charge states |0> and |1> are coupled by the Josephson energy, $E_J$, causing the bending of the dashed lines into the red solid curves. (**B**) A particular case of current flow via a TLS (see text). (**C**) Schematic evolution of the probability to be in state |1> as a function of time. The collapses to the |0> state occur when the TLS is emptied. For a superconducting charge qubit the oscillations have frequency $E_J/h$ and the collapse occurs by quasi-particle tunneling.

**Figure 2.** (**A**) Circuit with SIS detector capacitively coupled to a Josephson junction acting as a noise source. Both sides are connected to room temperature equipment via on-chip resistances, $R \approx 2$ k$\Omega$ made from platinium wires (dimensions: 20x0.1x0.02 $\mu m^3$). The on-chip capacitances, $C_c \approx 550$ fF, have dimensions: area is 80x10 $\mu m^2$ and silicon oxide insulator thickness is 50 nm. The circuit is mounted in a dilution refrigerator with base temperature 10 mK. (**B**) The black solid curve shows measured PAT current through the SIS detector for $V_{Jj} = 70$ $\mu$V. {Note $I_{PAT} \equiv I_{SIS}(V_{Jj}) - I_{SIS}(V_{Jj}=0)$.} For the red solid curve 2 the Josephson junction is DC current biased ($I_{Jj}=100$ nA) such that quasi-particle tunneling generates the noise. Eqn 1. is used to fit the dashed line to curve 1 obtaining $Z(\omega)$ and calculate without fit parameters the dashed-dotted curve.(Inset) Schematic energy diagram of PAT across an SIS junction. The curve is the bare $I_{SIS}$-$V_{SIS}$ characteristic (without noise) with the Josephson branch suppressed by a magnetic flux thanks to the SQUID geometry of the detector, $2\Delta = 420$ $\mu$eV. From the sample parameters we calculate the transimpedance $Z(\omega) \approx 900$ $\Omega$. This is an overestimation since it neglects the stray capacitances, for instance, from the Pt wires to ground.

**Figure 3.** (**A**) PAT current (logarithmic scale) versus $V_{Jj}$ and $V_{SIS}$. The right axis translates $V_{SIS}$ into frequency according to $f = (2\Delta - eV_{SIS})/h$. Zero current is black and high current is pink. The current scale can be inferred from curve 1 in Fig. 2B which is taken at the white dashed line. The yellow line separates the region $|V_{Jj}| < 2\Delta/e = 420$ $\mu$V, dominated by emission from the AC Josephson effect at frequency $2eV_{Jj}/h$ (indicated by the white line), and the region $|V_{Jj}| > 420$ $\mu$V where shot-noise is generated. The white area near $V_{Jj} = 0$ is not accessible due to the Josephson branch of the Josephson junction. The feature near $V_{SIS} = 0$ $\mu$V is due to the remaining Josephson branch of the SIS detector. (**B**) Transimpedance, $Z(\omega)$ versus frequency deduced from fitting the PAT current generated by the AC Josephson effect. (**C**) Shot-noise $S_I$ versus $I_{Jj}$. $S_I(I_{Jj})$ is deduced by scaling the dashed-dotted curve of Fig. 2B to fit the PAT current for different $I_{Jj}$. (For $I_{Jj}$ < 25 nA the Josepshon junction has not reached its normal state impedance, which causes the change in slope near the origin.)

**Figure 4. (A)** PAT current for different position on the JQP peak as indicated by arrows in the inset. The dashed line correspond to PAT current for $E_J/\Gamma$ < 1.(Inset) Current versus charge, $Q$, for the CBP on the JQP peak. **(B)** PAT current after subtracting the PAT curve for $Q/e$=1.3. The squares correspond to the $V_{SIS}$-value used for the dominant frequency determination. (Inset) Schematic picture of CPB. The superconducting island is connected to a superconducting lead via a Josephson junction with $E_J$ tuneable by a magnetic flux $\Phi$. A resistive junction ($R_P$ = 335 k$\Omega$) with negligible Josephson coupling is connected to the lower lead. A gate capacitance $C_g$ is used to modify the charge, $Q = C_g V_g$, on the island. **(C)** Dominant frequency deduced from the PAT current for two values of $E_J$. The solid lines are fits to the expression $\sqrt{(4E_C(Q/e-1))^2+E_J^2}/h$. The fit parameter $E_c$ = 100 µV is slightly higher than the measured charging energy. The arrows correspond to the points denoted on the left inset of (A). (Inset) Photo of CPB device with $E_c$= 95 µV deduced from Coulomb blockade experiments.

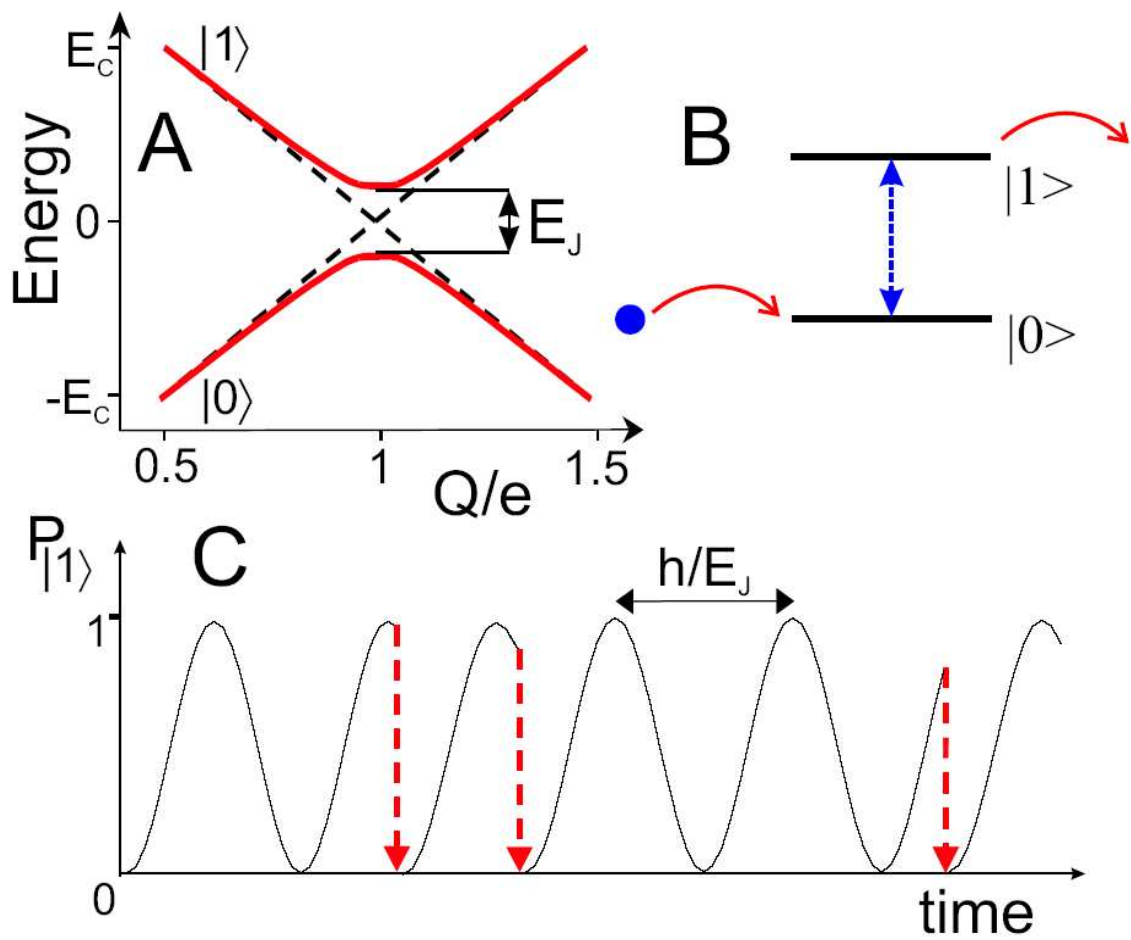

**Figure 1**

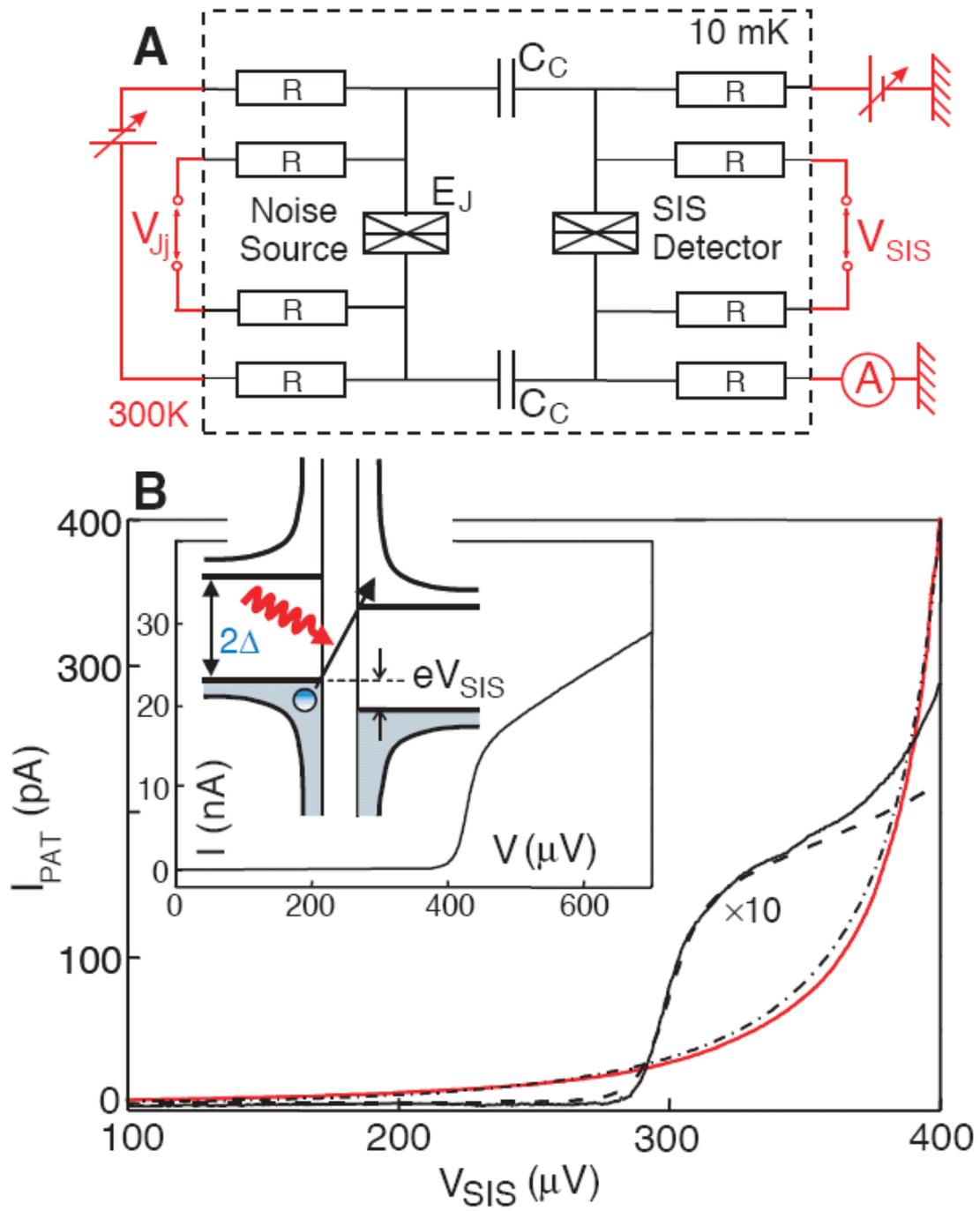

**Figure 2**

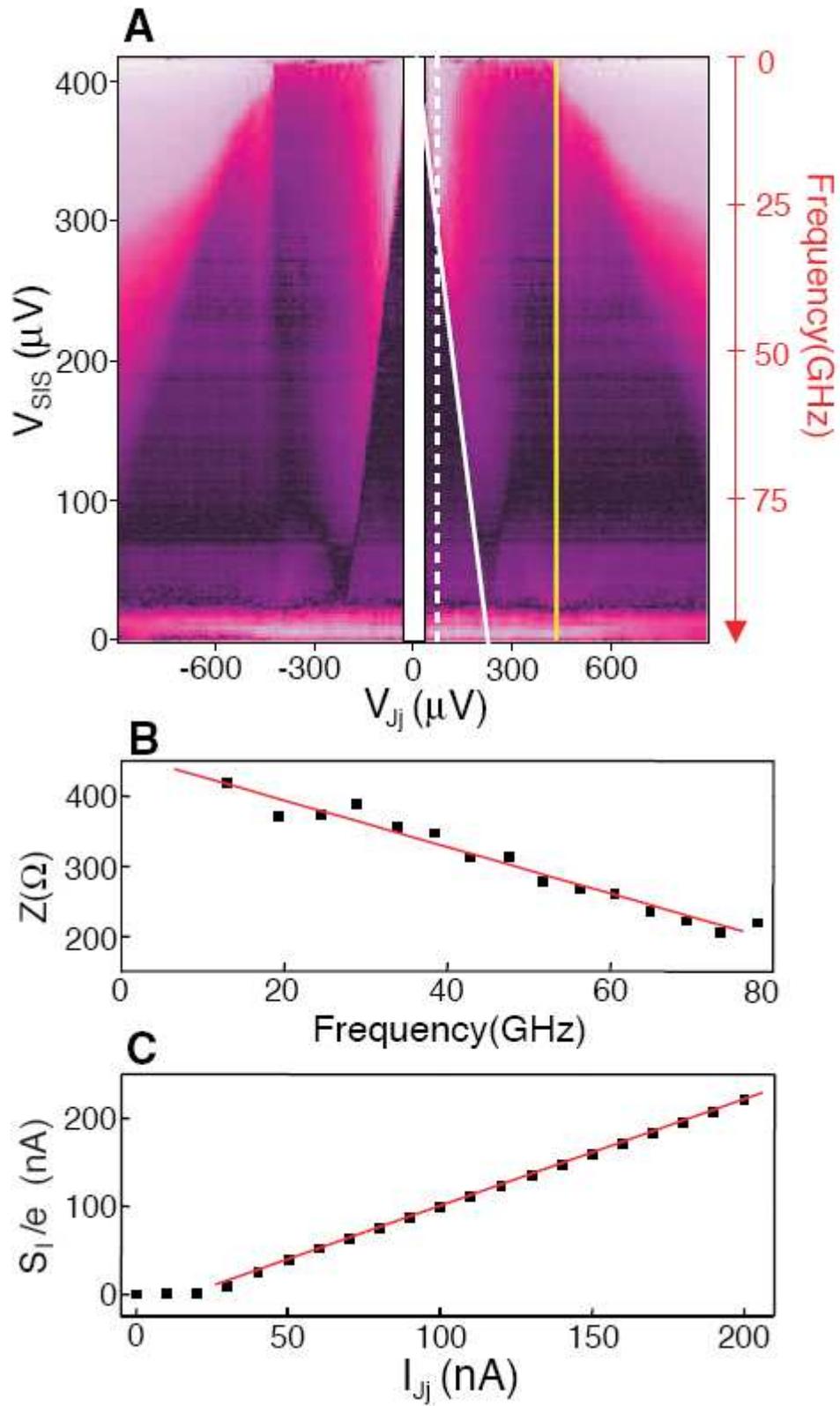

**Figure 3**

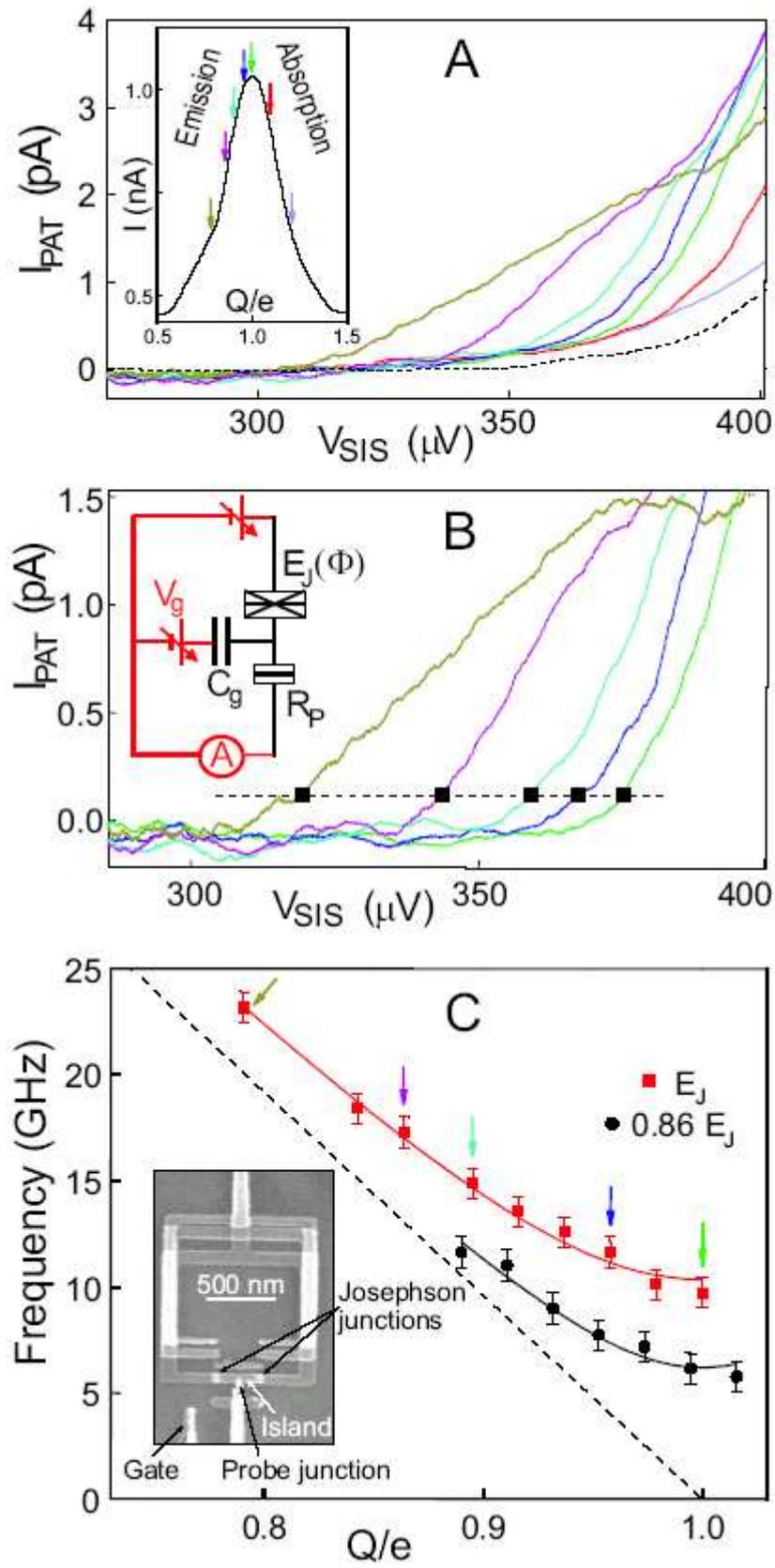

**Figure 4**

# Supporting Online Materials

## Methods and Fabrication

1- Fabrication

Both the sample with the Copper Pair Box (CPB) and the sample with the Josephson junction have the same on-chip circuitry. In the following we focus on the fabrication of the CPB (the source) coupled to the SIS detector (the detector).

In order to achieve a good coupling of the high frequency signal between the source and the detector we designed and fabricated an on-chip circuit comprising both of them. The AFM picture of the device is shown in Fig. S1.

We used a *Si* substrate with a 300nm *SiO* insulating layer on top. For the fabrication electron beam lithography and shadow evaporation technique was employed.

Three electron beam lithography steps were used. First one for the deposition of the on-chip resistances, a second one for the insulator layer of the capacitances and the third one for the source and detector themselves.

Both the detector and the source are provided with four contacts which allows us to measure the current and the real voltage drop at the same time.

The on-chip resistances are intended to prevent the leakage of the high frequency signal via stray-capacitances to ground. They were made out of a thin *Pt* layer with the following dimensions: 0.02 x 0.1 x 20 μm. Their measured resistance value was 3.3 kΩ at room temperature, respectively 1.9 kΩ at 20mK.

For coupling the source signal to the detector two big capacitances (10 X 80 μm) were fabricated. The bottom layer is made out of a 20 nm thick *Pt* layer; then a 50 nm thick *SiO* insulator layer and the top layer is made out of *Al* (120 nm thick). The estimated capacitance is 550 fF.

Both the SIS detector and the CPB are made out of *Al* and are in a SQUID geometry which allows us to tune the Josephson coupling by means of the magnetic field. Their area are in a ratio of 6:1 (1 x 1 μm and 2 x 3 μm) which allows for an independent tuning. The oxidation conditions were:
- for the detector and the Josephson junction of the CPB: 30 mT $O_2$ for 3 min,
- for the probe junction of the CPB: 45 mT $O_2$ with glow discharge for 5 min.

These resulted in resistances of $R_J$=16 kΩ respectively $R_P$=335 kΩ and capacitances of $C_J$=720 aF and $C_P$=121aF (see Fig. 4 of the article).

2- Power and efficiency

From the result on the sample with a Josephson junction coupled to a SIS detector we can evaluate the sensitivity of this noise detection scheme.

In the AC Josephson effect regime, the signal can be measured to frequencies as high as 90 GHz, for an estimated power emitted by the source junction of 100 fW. This power is converted into a PAT current with an efficiency of 0.03 at 25 GHz.

In the shot-noise regime the resolution is 80 fA$^2$/Hz, corresponding to an equivalent noise temperature of 3 mK on a 1 kΩ resistor.

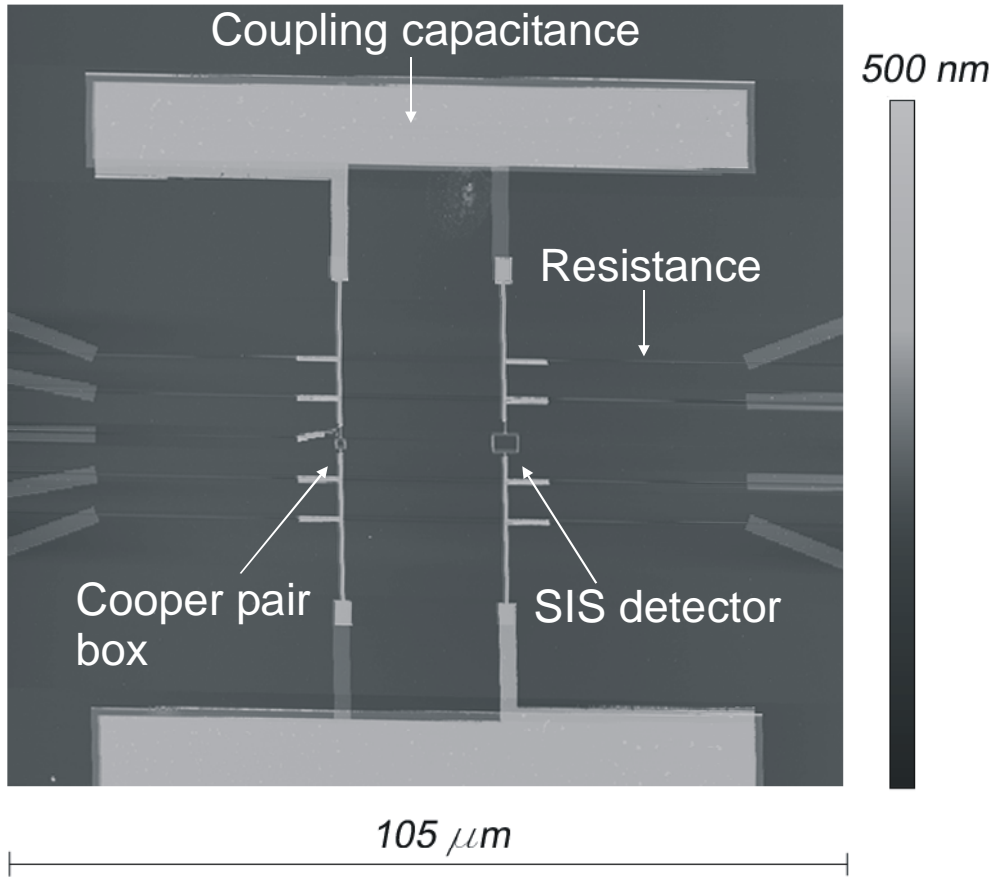

Fig. S1 AFM picture of the on-chip circuitry. The vertical bar on the right is a color scale for the height of the sample. The big pads of the coupling capacitors are visible at the top and the bottom of the picture. On the sides there are four contacts for the detector (on the right) and the source (on the left) plus one gate electrode for tuning the induced charge on the island of the CPB. Each contact is made through a *Pt* resistor (the thin part in the picture). The difference is area for the detector and the source is also clearly visible